\documentclass[aps,prd,superscriptaddress,twocolumn]{revtex4-2}
\usepackage{graphicx}
\usepackage[caption=false]{subfig}
\usepackage{epstopdf}
\usepackage{amsmath}
\usepackage{amsfonts}
\usepackage{amssymb}
\usepackage{latexsym}
\usepackage{hyperref}
\usepackage[english]{babel}
\usepackage[utf8]{inputenc}
\usepackage[colorinlistoftodos]{todonotes}
\usepackage{color}
\usepackage{slashed}
\usepackage{feynmp}
\usepackage{bm}
\usepackage{bbold}
\usepackage{eufrak}
\usepackage{slashed}
\usepackage{bm}  % To make greek letters in bold face {\bm \Phi}
\usepackage{tabu}
\usepackage{xcolor}
\begin{document}
	
	\title{Electroweak monopoles with a non-linearly realized weak hypercharge}

	\author{P. De Fabritiis}
	\email{pdf321@cbpf.br}
	\affiliation{Centro Brasileiro de Pesquisas F\'{i}sicas (CBPF), Rua Dr Xavier Sigaud 150, Urca, Rio de Janeiro, Brazil, CEP 22290-180}
	
	\author{J. A. Helayël-Neto}
	\email{helayel@cbpf.br}
	\affiliation{Centro Brasileiro de Pesquisas F\'{i}sicas (CBPF), Rua Dr Xavier Sigaud 150, Urca, Rio de Janeiro, Brazil, CEP 22290-180}

	\begin{abstract}
	
	We present a finite-energy electroweak-monopole solution obtained by considering non-linear extensions of the hypercharge sector of the Electroweak Theory, based on logarithmic and exponential versions of electrodynamics. We find constraints for a class of non-linear extensions and also work out an estimate for the monopole mass in this scenario. We finally derive a lower bound for the energy of the monopole and discuss the simpler case of a Dirac magnetic charge.
		
	\end{abstract}
	
	%%%%%%%%%%%%%%%%%%%%%%%%%%%%%%%%%%%%%%%%%%%	
	

	\maketitle

	\pagestyle{myheadings}

	%%%%%%%%%%%%%%%%%%%%%%%%%%%%%%%%%%%%%%%%%%%%

	\section{Introduction} \label{sec_intro}

``One would be surprised if Nature had made no use of it", said Dirac in 1931, talking about magnetic monopoles \cite{Dirac}. In this \textit{capolavoro}, Dirac shows that the existence of a magnetic monopole is not only consistent with the laws of Quantum Mechanics, but can provide an explanation for the quantization of electric charge and also render Maxwell equations more symmetric, realizing in an elegant way the duality symmetry. Since the groundbreaking work of Dirac, this fascinating subject has been theoretically explored in different circumstances, but despite the huge efforts to search for them experimentally, they remain undetected to this day.

In Dirac's work, it is not possible to predict what the monopole mass would be, since its classical energy is infinite by virtue of its singularity. Wu and Yang \cite{WuYang} generalized the concept for non-Abelian gauge theories showing that a pure SU(2) Yang-Mills theory also allows a point-like magnetic monopole, but also here the energy is infinite. 't Hooft and Polyakov \cite{Hooft, Polyakov} made a breakthrough discovery, finding a finite-energy monopole solution as a topological soliton in an SO(3) gauge theory with a scalar field in the adjoint representation, the so-called Georgi-Glashow model \cite{GG}. Here, for the first time, the monopole appears as a necessary prediction of the model instead of being only a consistent possibility and, remarkably, with a finite calculable mass.

Julia and Zee \cite{JuliaZee} extended the 't Hooft and Polyakov's solution by introducing a Coulombic part in the {\it ansatz}, and therefore finding a dyon solution, a particle with both electric and magnetic charges as introduced by Schwinger \cite{Schwinger}. Bogomol'nyi \cite{Bogomolnyi} and also Prasad and Sommerfield \cite{Prasad} found a special limit, nowadays called BPS limit, such that there is an analytical solution for the monopole (and dyon) and a lower bound for its energy. The monopole solution was constructed in Grand-Unified Theories by Dokos and Tomaras \cite{GUT}, and it's also relevant in the context of Supersymmetry and Dualities \cite{SusyDual}. 

The Electroweak (EW) Theory by Glashow, Salam and Weinberg \cite{EW} provides an extremely successful description for the unification of electromagnetic and weak interactions, and after the Higgs discovery in 2012 \cite{Higgs}, and all the others experimental tests in which it was successful, we can say that the Standard Model (SM) is in a very good shape. It's a very important question, therefore, to investigate if there exists an electroweak generalization of the 't Hooft-Polyakov monopole solution.

It was generally believed that such a solution would not be possible in the EW theory because the spontaneous symmetry breaking pattern of the EW gauge group $G = SU(2)_L \, \text{x} \, U(1)_Y \rightarrow H = U(1)_{em}$ does not allow a non-trivial second homotopy group, that is, we have $\pi_2 (G/H) = 0$. Nevertheless, there is an alternative topological scenario showing that the Standard Model admits an electroweak-monopole solution. In fact, it was originally shown by Cho and Maison \cite{CM} that, if we interpret the normalized Higgs doublet as a $CP^1$ field, we find the necessary topology to have a monopole solution since $\pi_2 (CP^1) = \pi_2 (S^2) = \mathbb{Z} $. It is sometimes said that this topologically stable EW monopole is somehow a non-trivial hybrid between the abelian Dirac monopole and the non-abelian 't Hooft-Polyakov monopole.

In their original work, Cho and Maison \cite{CM} present not only the topological scenario for the existence of electroweak monopoles, but also provide an explicit numerical solution for them, by assuming a spherically symmetric {\it ansatz}. The authors actually proved the existence of a more general electroweak dyon solution in the SM, and it is important to notice that an analytical existence theorem for such a solution can also be established \cite{Yang}. Unfortunately, this object suffers from a singularity in the origin, which yields an infinite energy at the classical level. A priori, there is nothing wrong with this, because the electron itself has an infinite electrostatic energy in Maxwell's Electrodynamics, though its mass is finite. This does not allow us to predict the mass of the monopole and, if we have a hope to find it experimentally, this becomes a non-trivial issue. Therefore, it is the purpose of this work to find a way to regularize the energy of the monopole solution and, then, infer about its mass.

There are already proposals of SM extensions giving regularized monopole solutions. One of them was proposed by Cho, Kimm, and Yoon \cite{CKY}, basically consisting in modifying the $U(1)_Y$ sector introducing a function depending on the magnitude of Higgs field in the usual hypercharge field strength  $- \frac{1}{4} \epsilon\left( \vert H \vert / v \right) B_{\mu \nu} B^{\mu \nu}$. It is possible to choose conditions on this $\epsilon$ such that we recover the usual SM in the standard electroweak vacuum and such that the energy integral is regularized at the origin, giving a finite-energy dyon solution. Roughly speaking they found a way to give an effective running to the $U(1)_Y$ coupling such that it can compensate the singularity present at the origin an therefore give a finite-energy dyon solution. The simple solution presented by the authors was latter shown by Ellis, Mavromatos and You \cite{Ellis1} to be incompatible with the LHC Data from the Higgs decay in two photons, but these authors were able to adjust their solution in a phenomenological consistent way, giving a family of possible solutions.

Following this line, Blaschke and Beneš \cite{BPSCM} were able to find a lower bound for the EW monopole mass by constructing a family of effective theories that have a BPS limit, in a way that the monopole mass can be found analytically and determined by the asymptotic behavior of the fields. Recently, Cho, Zhang and Zou \cite{Regcharge} shown that is possible to regularize the EW monopole energy by electric charge renormalization, founding a new BPS bound for the monopole solution.

Another interesting solution was proposed by Arunasalam and Kobakhidze \cite{Arunasalam}, where they considered an extension of the usual $U(1)_Y$ kinetic term to a non-linear Born-Infeld (BI) \cite{BI} Lagrangian. With this extension in the same way that the electron energy is regularized in the original BI Electrodynamics, the monopole energy here gets also regularized and its mass turns out to be proportional to the BI mass parameter $\beta$, that somehow controls the $U(1)_Y$ field non-linearity and can be constrained considering light-by-light scattering as was shown by Ellis, Mavromatos and You \cite{Ellis2}. The authors in \cite{Arunasalam} showed that a finite-energy monopole solution exists with this non-linear BI extension, and considered some consequences for the EW phase transition. Other interesting recent works are \cite{OutrosEW}.

The subject of non-linear extensions of Electrodynamics is a very rich research topic; it has been investigated in many different forms. There are very interesting features present in this type of theories, as for example the possibility of light-by-light scattering and the vacuum birefringence phenomenon. A very interesting property shared by some of them is the finite energy for the point charge, and this motivates us to investigate if this property can help us to regularize the infinite energy of the monopole solution in other non-linear extensions of the hypercharge sector. Therefore, in this contribution, we propose non-linear extensions of the hypercharge sector, following the extensions already inspected \cite{Helayel1, Helayel2} in the context of Electrodynamics, showing that it can lead to finite-energy monopole solutions; this allows us to estimate the mass of the monopoles as a function of the mass parameter associated to the non-linear extensions, in such a way that one can search experimentally for it in the MoEDAL \cite{MOEDAL} experiment at the LHC, for example. 

This work is organized as follows: in Section 2, we make a short review of the original EW monopole solution. In Section 3, we propose a regularization of the monopole solution, by adopting a non-linear extension of the hypercharge sector. Next, we analyze our results in some special cases, and estimate the respective masses. This is done in Section 4. Section 5 is devoted to finding a lower bound for the energy; finally, in Section 6, we cast our Concluding Remarks.

\section{The original electroweak monopole solution}
In this Section we do a brief review of the original Cho-Maison solution \cite{CM} and suggest the recent reviews \cite{review} for more details.

Let us consider the bosonic sector of the Electroweak Lagrangian in the Standard Model:
\begin{align}
	\mathcal{L}_0 = \vert D_\mu H \vert^2 - \frac{\lambda}{2} \left( H^\dagger H - \frac{\mu^2}{  \lambda} \right)^2 - \frac{1}{4} F_{\mu \nu }^a F^{\mu \nu }_a - \frac{1}{4} B_{\mu \nu } B^{\mu \nu },
\end{align}
where we are using $F_{\mu \nu }^a = \partial_\mu A_\nu^a - \partial_\nu A_\mu^a + g f^{a b c} A_\mu^b A_\nu^c
$ and $B_{\mu \nu } = \partial_\mu B_\nu - \partial_\nu B_\mu$. The covariant derivative with respect to the $SU(2)_L$ x $U(1)_Y$ gauge group is defined by, $D_\mu = \partial_\mu  - i \frac{g}{2} A_\mu^a \sigma^a   - i \frac{g'}{2}  B_\mu$,
%\begin{equation}\label{key}
%	D_\mu = \partial_\mu  - i \frac{g}{2} A_\mu^a \sigma^a   - i \frac{g'}{2}  B_\mu ,
%\end{equation} 
and $H$ is the SM Higgs doublet.
Given the above Lagrangian, one can obtain the equations of motion, given by
\begin{align}
	& D_\mu D^\mu H = - \lambda \left( H^\dagger H - \frac{\mu^2}{  \lambda} \right) H ; \nonumber \\ 
	& D_\mu F^{\mu \nu}_{a} = -i \frac{g}{2} \left(H^\dagger \sigma_a D^\nu H - D^\nu H^\dagger \sigma_a H \right) ; \nonumber\\
	& \partial_\mu B^{\mu \nu} = - i \frac{g'}{2} \left( H^\dagger D^\nu H - D^\nu H^\dagger H \right),
\end{align}
where we used $ D_\mu F_{\mu \nu}^{a} = \partial_\mu F_{\mu \nu}^a - g f^{a b c} A_\mu^c F_{\mu \nu}^b$.

Let us introduce here the following parametrization for the Higgs field, with no loss of generality,
\begin{align}\label{key}
	H = \frac{1}{\sqrt{2}} \, \rho \, \xi, \;\; \textrm{with} \;\; \xi^\dagger \xi = 1.
\end{align}
Here, the doublet structure is carried by the field $\xi$ as well as the Higgs hypercharge.
We would like to emphasize that the presence of the hypercharge quantum number in the field $\xi$ is extremely important to discuss the existence of the monopole solution. In fact, taking into account the $U(1)_Y$, we can interpret the unit doublet $\xi$ as a $CP^1$ field, and therefore find the non-trivial topology that we need to discuss monopole solutions. 

%Using this parametrization, and defining $\rho_0^2 = \frac{2 \mu^2}{\lambda}$, we can rewrite,
%\begin{align}
%	\mathcal{L}_0 &= \frac{1}{2} (\partial_\mu \rho)^2  + \frac{\rho^2}{2}  \vert D_\mu \xi \vert^2 - \frac{\lambda}{8} \left( \rho^2 - \rho_0^2 \right)^2 + \nonumber \\
%	& - \frac{1}{4} F_{\mu \nu }^a F^{\mu \nu }_a - \frac{1}{4} B_{\mu \nu } B^{\mu \nu }. 
%\end{align}
%The equations of motion written in terms of the new parametrization is,
%\begin{align}\label{key}
%	&\partial^2 \rho = \rho \vert D_\mu \xi \vert^2  - \frac{\lambda}{2} \left( \rho^2 - \rho_0^2 \right) \rho ; \\
%	&\rho D^2 \xi   = - 2 \, \partial_\mu \rho\, D_\mu \xi ;\\
%	&D_\mu F^{\mu \nu a} = -i \frac{g}{4} \rho^2 \left( \xi^\dagger \sigma^a D_\nu \xi - D_\nu\xi^\dagger \sigma^a \xi \right) ; \\
%	&\partial_\mu B_{\mu \nu} = -i \frac{g'}{4} \rho^2 \left( \xi^\dagger D_\nu \xi - D_\nu\xi^\dagger \xi \right).	
%\end{align}
Let us consider the spherically symmetric {\it ansatz}, proposed in \cite{CM}, with spherical coordinates $(t, r, \theta, \varphi)$:
\begin{align}\label{ansatz}
	&\rho = \rho(r), \quad \xi = i  \left( \begin{array}{c} \sin\left(\theta / 2\right) e^{- i \varphi} \nonumber \\ 
		- \cos\left( \theta/2 \right) 
	\end{array} \right) \nonumber \\
	&\vec{A}_\mu = \frac{1}{g} A(r) \, \partial_\mu t \; \hat{r} + \frac{1}{g}(f(r) - 1)\; \hat{r} \times \partial_\mu \hat{r}; \nonumber\\
	&B_\mu =  \frac{1}{g'} B(r) \, \partial_\mu t - \frac{1}{g'}(1 - \cos\theta) \, \partial_\mu \varphi.
\end{align}
Here we have $\hat{r} = - \xi^\dagger \, \vec{\sigma} \, \xi$, what in the abelian decomposition used in the original work would define the abelian direction in the gauge space. 
%Care should be paid to not confuse the coordinates in the field space with the spacetime coordinates in this spherically symmetric ansatz, where the interplay between both is important as usually happens in this subject. We also remark that again the $U(1)_Y$ is extremely important for these considerations, and without it the Higgs doublet would not allow a spherically symmetric ansatz, besides not having the correct topology to admit a monopole solution.
%Looking more closely to this ansatz, we can see that when we take $A(r) = B(r) = f(r) = 0$, we obtain a term like a magnetic potential for the field $\vec{A}_\mu$ that could describe a non-abelian monopole in the $SU(2)_L$ sector, and the field $B_\mu$ would give an abelian monopole in the $U(1)_Y$ sector. 
%There is an apparent string singularity in $\xi$ and $B_\mu$ along the negative z-axis, but this is only a gauge artifact, and can be removed. Therefore this ansatz is in the right direction to search for monopoles, and it is sometimes said that the Cho-Maison monopole is somehow an hybrid between the abelian Dirac monopole and the non-abelian 't Hooft-Polyakov monopole.  
Looking to this {\it ansatz} we can see that there is already a monopole structure both in $SU(2)_L$ and $U(1)_Y$ sectors.

Let us introduce the physical fields to understand better the content of the {\it ansatz}. To define the mass eigenstates in the gauge sector, we will choose the unitary gauge using a gauge transformation $U$ that puts the doublet in the usual form, that is $\xi \rightarrow (U \xi)^a = \delta^{a 2}$, with $a = 1 , 2 $.
%\begin{align}\label{key}
%	\xi \rightarrow U \xi =  \left( \begin{array}{c}
%		0 \\ 
%		1
%	\end{array} \right).
%\end{align} 
The transformation $U$ that does the job is
\begin{align}\label{key}
	U = i \left( \begin{array}{cc}
		\cos(\theta/2) & \sin(\theta/2) e^{-i \varphi} \\
		- \sin(\theta/2) e^{i \varphi} & \cos(\theta/2)
	\end{array} \right).
\end{align}
When we do such a gauge transformation, remembering that we have $\hat{r} = - \xi^\dagger \, \vec{\sigma} \, \xi$, we transform this abelian direction to $\hat{r}^a \rightarrow \delta^{a 3}$, with $a = 1, 2 , 3$.
%\begin{align}\label{key}
%	\hat{r} = \left( \begin{array}{c}
%		\sin\theta\cos\varphi\\
%		\sin\theta\sin\varphi\\
%		\cos\theta
%	\end{array} \right) \rightarrow \left( \begin{array}{c}
%		0\\
%		0\\
%		1
%	\end{array} \right).
%\end{align} 
Also the gauge fields have to change under this gauge transformation as usual, by $A_\mu' = U A_\mu U^{-1} - \frac{i}{g} \, \partial_\mu U \, U^{-1}$.
Therefore, in the unitary gauge, we have
\begin{align}\label{key}
	\vec{A}_\mu = \frac{1}{g} \left( \begin{array}{c}
		-f(r) \, (\sin\varphi \partial_\mu \theta + \sin\theta \cos\varphi \partial_\mu \varphi) \\
		f(r) \,  (\cos\varphi \partial_\mu \theta - \sin\theta \sin\varphi \partial_\mu \varphi)\\
		A(r) \partial_\mu t - (1 - \cos\theta) \partial_\mu \varphi
	\end{array} \right).
\end{align}
We define the physical fields $A_\mu$ and $Z_\mu$ through the rotation with the Weinberg angle, that is, $Z_\mu = \cos\theta_W A_\mu^3 - \sin\theta_W B_\mu$ and $A_\mu = \sin\theta_W A_\mu^3 + \cos\theta_W B_\mu $,
%\begin{align}\label{key}
%	\left( \begin{array}{c}
%		Z_\mu \\
%		A_\mu
%	\end{array} \right) = \left( \begin{array}{cc}
%		\cos\theta_W & -\sin\theta_W\\
%		\sin\theta_W & \cos\theta_W
%	\end{array} \right) 	\left( \begin{array}{c}
%		A_\mu^3 \\
%		B_\mu 
%	\end{array} \right),
%\end{align}
and we define also the W-bosons through $W^{\pm}_\mu = \frac{1}{\sqrt{2}}\left( A_\mu^1 \mp i A_\mu^2 \right)$. Plugging the {\it ansatz}, we obtain
\begin{align}\label{key}
	&A_\mu^{\text{em}} = e \left( \frac{1}{g^2} A(r) + \frac{1}{g'^2} B(r) \right) \partial_\mu t - \frac{1}{e}(1 - \cos\theta) \partial_\mu\varphi ;\nonumber\\
	&Z_\mu = \frac{e}{g g'} (A(r) - B(r)) \partial_\mu t ;\nonumber\\
	&W_\mu^- = \frac{i}{g} \frac{f(r)}{\sqrt{2}} e^{i \varphi} (\partial_\mu \theta + i \sin\theta \partial_\mu \varphi),
\end{align}
%where we defined $e = g \sin\theta_W = g' \cos\theta_W$. %Therefore, we have our ansatz written in terms of the physical fields, and we can begin to search for solutions of the equations of motion.

The equations of motion will give us a set of coupled differential equations in the radial variable for the fields $(A(r), B(r), f(r), \rho(r))$. 
%Notice that we will consider that all objects are constant in time since we are searching for static solutions, 
We point out that we are considering here only static solutions, and therefore, the derivatives are taken with respect to the variable $r$. The spherical symmetry of the {\it ansatz} simplifies considerably the equations of motion, and one can show that these equations admit a general dyon solution if we impose certain boundary conditions \cite{CM}.
The energy-momentum tensor here is given by
\begin{align}\label{tensor0}
	T^{\mu \nu}_0 = &F^{\mu \rho}_a \, F^{a \, \nu }_\rho + B^{\mu \rho} B_\rho^{\,\,\, \nu} - \eta^{\mu \nu} \mathcal{L}_0 \; + \nonumber \\
	&+  D^{\mu} H^\dagger D^\nu H + D^{\nu} H^\dagger D^\mu H .
\end{align}
Therefore, the energy functional for the {\it ansatz} \ref{ansatz} is
%From here, one we can obtain the energy functional and evaluate this expression in the configurations given by the ansatz, to obtain the energy associated with this solution. Therefore, we can obtain:
\begin{align}\label{key}
	&E = 4 \pi \int_0^\infty dr \,  r^2 \left[ \frac{\rho^2}{8} (A- B)^2 + \frac{\dot{\rho}^2}{2} + \frac{\rho^2 f^2}{4}\frac{1}{r^2} \right.    \nonumber \\
	&+  \left.  \frac{\dot{A}^2}{2g^2}   + \frac{A^2 f^2}{g^2} \frac{1}{r^2} + \frac{\dot{f}^2}{g^2} \frac{1}{r^2} + \frac{(f^2-1)^2}{2 g^2} \frac{1}{r^4}  \right.  \nonumber \\ 
	& + \left. \frac{\lambda}{8}(\rho^2 - \frac{2 \mu^2}{\lambda})^2    +    \frac{\dot{B}^2}{2 g'^2} + \frac{1}{2 g'^2} \frac{1}{r^4}      \right] .
\end{align}
%From this expression, using the appropriate boundary conditions for the functions $ \rho(r), f(r), A(r), B(r)$, one can show that most of the terms above give a finite result, but the last term in the above expression will become infinite, and we will call this contribution  $E^*$, that is,
We remark that in this expression all the terms give a finite contribution except the last one, that we will call
\begin{align}\label{key}
	E^* = 4 \pi \int_0^\infty dr \, \frac{1}{2 \, g'^2 \, r^2} .
\end{align}
This is exactly the origin of the infinite energy of the monopole solution at the classical level, a singularity at the origin. Because of this problem, we cannot predict the monopole mass, and we will propose in the following a solution to this issue, finding a finite energy monopole. 
%it is important to note that these are classical considerations, and one can consider extensions of the Standard Model that can regularize this divergence, paying special attention to the $U(1)_Y$ sector where it appears.

\section{A finite-energy monopole solution}
%Here we are interested in the simpler case of a monopole solution, since it is lighter than the dyon and more easily accessible in a experimental sense. Therefore we will take in our analysis the simplifying assumption $A(r) = B(r) = 0$, that is, we will turn of the Coulombic part of the Ansatz, to investigate the electroweak monopole solutions. 
Let us consider now the simpler case of a monopole solution, since it is lighter than the dyon and more easily accessible in an experimental sense. That is, we will consider the simplified version of the more general {\it ansatz} \ref{ansatz} where we turn off the Coulombic part taking $A(r) = B(r) = 0$. The EW monopole {\it ansatz } is therefore given by
%the ansatz in the particular case of $A = B = 0$, that is, restricting attention to the monopole solution and ignoring the possibility of dyons, for the reasons that we already explained before. Therefore, repeating the important expressions here for the comfort of the reader, the restricted ansatz gives,
\begin{align}\label{monopole ansatz}
	&\rho = \rho(r), \quad \xi = i  \left( \begin{array}{c} \sin\left(\theta / 2\right) e^{- i \varphi} \nonumber\\ 
		- \cos\left( \theta/2 \right) 
	\end{array} \right) \nonumber\\
	&\vec{A}_\mu = \frac{1}{g}(f(r) - 1)\; \hat{r}  \times  \partial_\mu \hat{r}; \nonumber \\
	&B_\mu =  - \frac{1}{g'}(1 - \cos\theta) \, \partial_\mu \varphi.
\end{align}
%The Lagrangian in this particular case reduces to,
%\begin{align}\label{key}
%	L = 4 \pi \int_0^\infty dr \, r ^2 \left[\left( \frac{\dot{\rho}^2}{2} +    \frac{\rho^2 f^2}{4}\frac{1}{r^2} - \frac{\lambda}{8}\left( \rho^2 - \rho_0^2 \right)^2\right) + \left(  - \frac{\dot{f}^2}{g^2} \frac{1}{r^2} - \frac{(f^2-1)^2}{2 g^2} \frac{1}{r^4}  \right) + \left(    - \frac{1}{2 g'^2}\frac{1}{r^4}\right) \right].
%\end{align}
The equations of motion are simplified in this case to
\begin{align}\label{key}
	&\ddot{\rho} + \frac{2}{r} \dot{\rho} - \frac{f^2}{2 r^2} \rho =  \frac{\lambda}{2} \left( \rho^2 - \frac{2 \mu^2}{\lambda} \right) \rho ;\nonumber\\
	&\ddot{f} - \frac{(f^2 - 1)}{r^2} f =  \frac{f g^2}{4} \rho^2 .
\end{align} 
The monopole {\it ansatz} \ref{monopole ansatz} provides a solution to the equations of motion if we adopt the following boundary conditions \cite{CM}:
\begin{align}\label{asymptotic}
	\rho(0) = 0 , \, \rho(\infty) = \rho_0, \, f(0) = 1, \, f(\infty) = 0,
\end{align}
where we defined $\rho_0 = \sqrt{\frac{2 \mu^2}{\lambda}}$. The energy functional for the monopole configuration is also simplified, giving us
\begin{align}\label{monopole energy}
	E &= 4 \pi \int_0^\infty dr \,  r^2 \left[  \frac{\dot{\rho}^2}{2} + \frac{\rho^2 f^2}{4}\frac{1}{r^2} + \frac{\lambda}{8}(\rho^2 - \rho^2_0)^2 \right.  \nonumber \\  
	&+ \left.    \frac{\dot{f}^2}{g^2} \frac{1}{r^2} + \frac{(f^2-1)^2}{2 g^2} \frac{1}{r^4}    +   \frac{1}{2 g'^2} \frac{1}{r^4}      \right], 
\end{align}
such that one can write for simplicity in the following,
\begin{equation}\label{key}
	E = E_1 + E^*.
\end{equation}
We remark that the problematic term $E^*$ is still here, as one can see in the last term of the above expression. 
%But in this particular case, it is easier to understand the system and handle the equations, and as the monopole is lighter than the general dyon, its where we will focus and try to search for solutions to the infinite energy problem.

Let us propose the following general extension of $U(1)_Y$ sector in the EW Lagrangian $\mathcal{L}_0$, considering:
\begin{align}\label{key}
	\mathcal{L} = \mathcal{L}_0 + f\left( \mathcal{F} , \mathcal{G} \right),
\end{align}
where we define $\mathcal{F} = \frac{1}{4} B_{\mu \nu} B^{\mu \nu}$, and $\mathcal{G} = \frac{1}{4} B_{\mu \nu} \tilde{B}^{\mu \nu} $, the $U(1)_Y$ Lorentz and gauge invariant basic objects, where $\tilde{B}^{\mu \nu} = \frac{1}{2} \epsilon^{\mu \nu \rho \sigma} B_{\rho \sigma}$.
In this case, the equations of motion for the $U(1)_Y$ sector will become
\begin{align}\label{key}
	\partial_\mu B^{\mu \nu} = \frac{ J^\nu + B^{\mu \nu } \partial_\mu \partial_\mathcal{F} f + \tilde{B}^{\mu \nu} \partial_\mu \partial_\mathcal{G} f + \partial_\mu \tilde{B}^{\mu \nu} \partial_\mathcal{G} f }{1 - \partial_\mathcal{F} f} ,
\end{align}
where we defined $\partial_\mathcal{F} f = \frac{\partial f}{\partial \mathcal{F} }$, $\partial_\mathcal{G} f = \frac{\partial f}{\partial \mathcal{G} }$, and we defined also the hypercharge matter current $J^{\nu} = -i \frac{g'}{2}  \left( H^\dagger D^\nu H - D^\nu H^\dagger H \right) $.
Now, we can plug the {\it ansatz} in this equation of motion to see which constraints we obtain. Notice that here we have $\partial_0 = 0$, and $B_0 = 0$, and thus immediately we obtain $B_{i 0} = 0 $, $\tilde{B}_{i j } = 0$. After some algebraic manipulations, we can also obtain $J^\nu = 0$. 
We can write $B^{i j} \propto \epsilon^{i j k} \mathcal{B}_k(r) $, where $\vec{\mathcal{B}}(r)$ is the radial hypercharge magnetic field associated with the $U(1)_Y$ gauge potential, and thus one can find $\partial_i B^{i j} = 0$. 
%We can also obtain,
%\begin{align}\label{key}
%	\left( \xi^\dagger D_\mu \xi - D_\mu\xi^\dagger \xi \right) &= i \left[ g A_\mu^1 \sin\theta \cos\varphi + g A_\mu^2 \sin\theta \sin\varphi \right. \nonumber \\
%	&\left. + g A_\mu^3 \cos\theta - g' B_\mu - (1 - \cos\theta) \partial_\mu \varphi  \right] \nonumber \\
%	&= 0.
%\end{align}
The equation of motion, after these considerations, can be written as
%\begin{align}\label{key}
%	 \partial_i B^{i \nu} = \frac{B^{i \nu} \partial_i (\partial_\mathcal{F}f) +  \tilde{B}^{i \nu} \partial_i (\partial_\mathcal{G}f) + \partial_i \tilde{B}^{i \nu} (\partial_\mathcal{G} f)}{1 - \partial_\mathcal{F}f}.
%\end{align}
%\begin{align}\label{key}
%	\partial_i B^{i \nu} -  \partial_i \left( B^{i \nu}  \partial_\mathcal{F} f \right)  -  \partial_i \left( \tilde{B}^{i \nu}  \partial_\mathcal{G} f \right)   = 0.
%\end{align}
\begin{align}\label{key}
	B^{i \nu} \partial_i  \partial_\mathcal{F} f   + \tilde{B}^{i \nu} \partial_i  \partial_\mathcal{G} f  + \partial_i \tilde{B}^{i \nu} \partial_\mathcal{G} f  = 0.
\end{align}
%But we have $B_{i j} = - \frac{1}{g'} \sin\theta \left( \partial_i\theta \, \partial_j\varphi - \partial_j\theta \, \partial_i\varphi \right)$, and 
%thus we can also obtain $\partial_i B^{i j } = 0$. 
%After a few manipulations, we can obtain the following equation,
%\begin{align}\label{key}
%	B^{i \nu} \partial_i \left( \partial_\mathcal{F} f \right) + \tilde{B}^{i \nu} \partial_i \left( \partial_\mathcal{G} f \right) + \partial_i \tilde{B}^{i \nu} (\partial_\mathcal{G} f)  = 0.
%\end{align}
Therefore, given our proposal of extending the hypercharge sector adding a generic function $f\left(\mathcal{F}, \mathcal{G} \right)$, we conclude that the monopole {\it ansatz} will satisfy the modified $U(1)_Y$ equation of motion if the function $f(\mathcal{F} , \mathcal{G})$ satisfies the following conditions:
\begin{align}\label{solutioncondition}
	\partial_i  \left( B^{i j}  \partial_\mathcal{F} f \right)\vert_{ansatz} = 0, \nonumber \\
	\partial_i \left( \tilde{B}^{i 0}  \partial_\mathcal{G} f \right)  \vert_{ansatz} = 0.
\end{align}

Now, let us study the energy of the monopole configuration in this extended model. One can obtain the following energy-momentum tensor:
\begin{align}\label{key}
	\tilde{T}^{\mu \nu } &= T^{\mu \nu}_0   + B^{\mu \rho } \partial^\nu B_\rho \, \partial_\mathcal{F} f + \tilde{B}^{\mu \rho} \partial^\nu B_\rho \, \partial_\mathcal{G} f - \eta^{\mu \nu } f + \nonumber \\
	& - \left(\frac{ J^\mu + B^{\rho \mu } \partial_\rho \partial_\mathcal{F} f  + \partial_\rho \left(\tilde{B}^{\rho \mu} \partial_\mathcal{G} f\right) }{1 - \partial_\mathcal{F} f}\right) B^\nu + J^\mu B^\nu, 
\end{align}
where $T^{\mu \nu}_0$ is the usual energy-momentum tensor \ref{tensor0}. 
Now we can take the Hamiltonian and calculate the monopole configuration energy, simply plugging our {\it ansatz} into this expression. In the monopole {\it ansatz}, we remember again that $\partial_0 = 0$ and $B_0 = 0$, giving us a huge simplification. In fact, since we have $B_{0 i} = \tilde{B}_{i j} = 0$, we can immediately obtain $B_{\mu \nu} B^{\mu \nu}\propto \vec{\mathcal{B}}(r)^2$ and also $B_{\mu \nu} \tilde{B}^{\mu \nu} = 0$, giving us  
\begin{align}\label{FGansatz}
	&\mathcal{F}\vert_{ansatz} = \frac{1}{2 g'^2 }\frac{1}{r^4}, \nonumber \\
	&\mathcal{G}\vert_{ansatz} = 0.
\end{align}
 Thus, the monopole energy in this extended model is
\begin{align}\label{monopole energy}
	E = E_1 +  \int_0^\infty dr 4 \pi r^2 \left[ \frac{1}{2 g'^2}\frac{1}{r^4} - f\left( \mathcal{F}\vert_{ansatz} ; \mathcal{G}\vert_{ansatz} \right) \right].
\end{align}
%With no loss of generality, we can define for convenience,
%\begin{align}\label{key}
%	f\left( \mathcal{F} ; \mathcal{G} \right) = \mathcal{F} + \phi \left( \mathcal{F} , \mathcal{G} \right).
%\end{align}
%Summarizing, we conclude that considering the following $U(Y)_Y$ extended model,
%\begin{align}\label{key}
%	\mathcal{L} = \vert D_\mu H \vert^2 - \frac{\lambda}{2} \left( H^\dagger H - \frac{\mu^2}{  \lambda} \right)^2 - \frac{1}{4} F_{\mu \nu }^a F^{\mu \nu }_a + \phi\left( \mathcal{F} , \mathcal{G} \right),
%\end{align}
%where $\phi\left( \mathcal{F} , \mathcal{G} \right)$ is a generic function of the invariants $\mathcal{F} = \frac{1}{4} B_{\mu \nu} B^{\mu \nu}$, and $\mathcal{G} = \frac{1}{4} B_{\mu \nu} \tilde{B}^{\mu \nu}$, the monopole ansatz is a solution of the equations of motion if $\phi$ satisfies the following conditions
%\begin{align}\label{key}
%	B^{i j} \partial_i \left( \partial_\mathcal{F} \phi \right)\vert_{ansatz} = 0, \\
%	\tilde{B}^{i 0} \partial_i \left( \partial_\mathcal{G} \phi \right)\vert_{ansatz} = 0,
%\end{align}    
%where $\mathcal{F}\vert_{ansatz} = \frac{1}{2 g'^2 }\frac{1}{r^4}$ and $\mathcal{G}\vert_{ansatz} = 0$, and this monopole configuration will have finite energy if it satisfies the following condition,
%\begin{align}\label{key}
%	- \int_0^\infty dr 4 \pi r^2 \left[   \phi \left( \mathcal{F}\vert_{ansatz} ; \mathcal{G}\vert_{ansatz} \right) \right] = \text{Finite}.
%\end{align}
Notice that we can easily handle the infinite energy coming from $E^*$ simply taking  $f\left( \mathcal{F} , \mathcal{G} \right) = \mathcal{F} + \phi \left( \mathcal{F} , \mathcal{G} \right) $.
Therefore, we can use the expressions \ref{solutioncondition} and \ref{monopole energy} to search for extensions of the $U(1)_Y$ sector of the electroweak Lagrangian such that the monopole {\it ansatz} is a finite energy solution for the equation of motion.

In fact, let us impose that this function $\phi \left( \mathcal{F} , \mathcal{G} \right)$, that will represent our generalized $U(1)_Y$ kinetic term, depends non-trivially on $\mathcal{F}$ and only on the square of $\mathcal{G} = \frac{1}{4}B_{\mu \nu}\tilde{B}^{\mu \nu}$, that is, $\phi$ is a generic function of $\mathcal{F}$ and $\mathcal{G}^2$. The physical reason for this assumption is to not have a parity violating term in the gauge kinetic sector of the photon after the EW symmetry breaking. 
One can show that, only imposing this physical assumption, the conditions \ref{solutioncondition} will be trivially satisfied for any reasonable function $\phi$, that is, the monopole {\it ansatz} will satisfy the equation of motion coming from the extended $U(1)_Y$ sector. We remark here that this is a sufficient condition to solve the constraints \ref{solutioncondition}, but it is not necessary. 

Therefore, the most general extension of the hypercharge sector for which the monopole {\it ansatz} \ref{monopole ansatz} is a solution of the equations of motion, and consistent with the above physical assumption is any reasonable function $\phi\left( \mathcal{F} , \mathcal{G}^2 \right) $ such that the energy integral is finite, {\it i.e.},
\begin{align}\label{key}
	- \int_0^\infty dr 4 \pi r^2 \left[ \phi \left( \mathcal{F} = \frac{1}{2 g'^2 r^4} ; \mathcal{G}^2 = 0 \right) \right]	 = \text{Finite}
\end{align}

In particular, since we want to reproduce the usual $- \frac{1}{4} B_{\mu \nu}B^{\mu \nu}$ term in first approximation to recover the SM results at first order, we will study a restricted class of possible extensions considering that $\phi$ depends on $\mathcal{F}$ and $\mathcal{G}$ through the particular combination $X = \frac{\mathcal{F}}{\beta^2} -\frac{\mathcal{G}^2}{2 \beta^4} $, where $\beta$ is a parameter with dimensions of $\text{Mass}^2$. As we already know, the conditions \ref{solutioncondition} are trivially satisfied, and we need only to care about the finiteness of the energy integral. What we are doing here is to improve the hypercharge sector to a non-linear version, and we will consider three physically interesting cases, corresponding to $\phi_1 = - \beta^2 \log\left[1 + X \right] $, $\phi_2 = \beta^2 \left[ e^{-X } -1 \right] $ and finally, $\phi_3 = \beta^2\left[ 1 - \sqrt{1 + 2 X }\right] $, that respectively will give us the $U(1)_Y$ version of the Logarithmic \cite{Helayel1}, Exponential \cite{Helayel2}, and Born-Infeld \cite{BI} non-linear Electrodynamics.

\section{Non-linear extensions of $U(1)_Y$}

The subject of non-linear Electrodynamics was introduced in the thirties by Euler and Heisenberg \cite{EulerHeisenberg} after the Nature's paper by Born and Infeld \cite{BI} to remove the singularities associated with charged point-like particles, and it has ever since attracted the interest of physicists due to its interesting features. For example, non-linear Electrodynamics predicts light-by-light scattering in vacuum and such phenomenon is being tested experimentally nowadays \cite{Light}. Interestingly, some non-linear models emerge naturally from the low-energy limit of string theory, and it has been applied in very different contexts as, for example, black hole physics \cite{BH}, holographic superconductivity \cite{superconductor}, and cosmology \cite{cosmology}. There are, nowadays, many different proposals of non-linear Electrodynamics \cite{NonLinear}, exhibiting not only finite energy for the point-like charge, but also properties like vacuum birefringence and dichroism.

In this Section, we shall consider three possible non-linear extensions of the hypercharge sector, calculate the monopole energy for each of them, and compare the respective results. We remark here that the Born-Infeld case was already studied in \cite{Arunasalam}, and we are considering these results here only for the sake of comparison.  In each of the following cases, what we will do is to consider different functions $\phi = \mathcal{L}_Y$, that extends the hypercharge sector to a non-linear theory, state its equation of motion, and compute the corresponding monopole energy integral for it. The right-hand side of the equation of motion will be given by the usual matter current $J^\nu = - \frac{i g'}{2} \left( H^\dagger D^\nu H - D^\nu H^\dagger H \right) $. All  of them have a factor $E_1$ in common, since this is the contribution to the energy that comes from all the other terms except the $U(1)_Y$ kinetic term. As we already remarked before, this contribution $E_1$ is finite, and its value was calculated by \cite{CKY}, giving approximately $ E_1 \approx 4.1 \, \text{TeV}$.

Let us consider first the Logarithmic $U(1)_Y$ Electrodynamics, introduced few years ago \cite{Helayel1}. The Lagrangian for the hypercharge sector will be
\begin{align}\label{key}
\mathcal{L}_{Y} = -\beta^2 \log\left[ 1 + \frac{\mathcal{F}}{\beta^2} - \frac{\mathcal{G}^2}{2 \beta^4} \right],
\end{align}
where as before, $\mathcal{F} = \frac{1}{4} B_{\mu \nu}B^{\mu \nu}$ and $\mathcal{G} = \frac{1}{4} B_{\mu \nu}\tilde{B}^{\mu \nu} $, and $\beta$ is a parameter with dimensions of $ \text{Mass}^2$. The $U(1)_Y$ equation of motion for our extended theory is
\begin{align}\label{key}
	\partial_\mu \left[ \frac{B^{\mu \nu} - \frac{1}{\beta^2} \mathcal{G} \tilde{B}^{\mu \nu}}{ \left(1 + \frac{\mathcal{F}}{\beta^2} - \frac{\mathcal{G}^2}{2 \beta^4}\right)} \right] = J^\nu,
\end{align}
The monopole energy here is given by
\begin{align}\label{key}
	E = E_1 +  \int_0^\infty dr  \left[ \beta^2 \log\left( 1 + \frac{1}{2 \beta^2 g'^2 r^4} \right) 4 \pi r^2	\right].
\end{align}
Doing this integral we obtain
\begin{align}\label{key}
	E = E_1 + \frac{2}{3} 2^{3/4} \pi^2 \frac{\sqrt{\beta}}{(g')^{3/2}}.
\end{align}
To estimate the energy, we will consider here $g' = 0.357$, that is approximately the value of the $U(1)_Y$ coupling at the EW scale. Thus, we obtain
\begin{align}\label{key}
	E \approx 4.1 \, \text{TeV} + 51.87 \sqrt{\beta}.
\end{align}

Now, let us consider the Exponential $U(1)_Y$ Electrodynamics \cite{Helayel2}. Here we have the following Lagrangian:
\begin{align}\label{key}
	\mathcal{L}_{Y}  = \beta^2 \left[ -1 + \exp^{\left( - \frac{\mathcal{F}}{\beta^2} + \frac{\mathcal{G}^2}{2 \beta^4} \right)}  \right].
\end{align}
The equation of motion follows immediately,
\begin{align}\label{key}
	\partial_\mu \left[ \left( B^{\mu \nu} - \frac{1}{\beta^2} \mathcal{G} \tilde{B}^{\mu \nu} \right) \exp^{\left( -\frac{\mathcal{F}}{\beta^2} + \frac{\mathcal{G}^2}{2 \beta^4} \right)} \right]  = J^\nu.
\end{align}
Repeating the same steps, we can find the energy integral,
\begin{align}\label{key}
	E = E_1 + \int_0^\infty dr 4 \pi r^2 \beta^2 \left[ 1  -   \exp^{\left( - \frac{\mathcal{F}}{\beta^2} + \frac{\mathcal{G}^2}{2 \beta^4} \right)}  \right].
\end{align}
Doing this integral, we obtain
\begin{align}\label{key}
	E = E_1 - \frac{\pi}{2^{3/4}} \Gamma(-3/4) \frac{\sqrt{\beta}}{(g')^{3/2}}.
\end{align}
Using as before $g' = 0.357$, we obtain
\begin{align}\label{key}
	E \approx 4.1 \, \text{TeV} + 42.33 \, \sqrt{\beta}.
\end{align}

Last, but not least, we introduce the well-known Born-Infeld case, that have the following Lagrangian:
\begin{align}\label{key}
	\mathcal{L}_{Y} = \beta^2 \left[ 1 - \sqrt{1 + \frac{2}{\beta^2} \mathcal{F} - \frac{1}{\beta^4}\mathcal{G}^2}  \right].
\end{align}
The equation of motion here is
\begin{align}\label{key}
	\partial_\mu \left[ \frac{B^{\mu \nu}  - \frac{1}{ \beta^2}\mathcal{G} \tilde{B}^{\mu \nu}    }{\sqrt{1 + \frac{2}{\beta^2} \mathcal{F}- \frac{1}{\beta^4} \mathcal{G}^2}}   \right] = J^\nu,
\end{align}
and the energy integral is given by
\begin{align}\label{key}
	E = E_1 + \int_0^\infty dr 4 \pi r^2 (- \beta^2) \left[ 1 - \sqrt{1 + \frac{2}{\beta^2} \mathcal{F} - \frac{1}{\beta^4}\mathcal{G}^2}  \right].
\end{align}	
Solving this integral, we obtain
\begin{align}\label{key}
	E = E_1 + \frac{3 \sqrt{\pi} \, \Gamma(-3/4)^2}{8 \, (g')^{3/4}} \sqrt{\beta}
\end{align}
Taking $g' = 0.357$, we have
\begin{align}\label{key}
	E \approx 4.1 \, \text{TeV} + 72.81 \sqrt{\beta}.
\end{align}

Now that we already have the expressions for the energy, let us discuss a little bit these $U(1)_Y$ extensions. First of all, we can see that if we perform a Taylor expansion of them in the parameter $1/\beta^2$, we obtain at first non-trivial order,
\begin{align}\label{key}
	\mathcal{L}_Y = -\mathcal{F} +  \frac{1}{2 \beta^2} \left[ \mathcal{F}^2 + \mathcal{G}^2 \right]  + O\left(1/\beta^4\right).
	%&= - \frac{1}{4} B_{\mu \nu} B^{\mu \nu} \! + \! \frac{1}{2 \beta^2} \left[ \!\left( \frac{1}{4} B_{\mu \nu} B^{\mu \nu} \right)^2 \!\!\! + \! \left( \frac{1}{4} B_{\mu \nu} \tilde{B}^{\mu \nu} \right)^2 \! \right]
\end{align}
Notice that they reproduce the usual kinetic term at first order, and exactly agree at order $O(1/\beta^2)$. This $\sqrt{\beta}$ parameter with dimensions of energy controls somehow the non-linearity of the fields,  and can be obtained from experiments, but we notice that it should be large in comparison to our scales of energy since we do not observe non-linear effects at low energy. The best known bound for the $\beta $ parameter nowadays is given by the work  \cite{Ellis2} considering Data from light-by-light scattering measurements in LHC Pb-Pb collisions by ATLAS, and gives a lower bound for the Born-Infeld parameter in Electrodynamics given by $\sqrt{\beta} \ge 100 \, \text{GeV}$. Here, we are doing a non-linear extension in the hypercharge sector instead of directly in the Electrodynamics, therefore, we should take a factor of $\cos\theta_W$ into account, obtaining the bound $\sqrt{\beta} \ge 90 \, \text{GeV}$. In principle, one should take for each non-linear $U(1)_Y$ extension a different bound for the corresponding $\beta$ parameter, but we can argue that we can consider all of them approximately equal with a good approximation. In fact, as the bound was obtained considering light-by-light scattering, the relevant term is the one with 4 photons in it, coming from the terms $\left(F_{\mu \nu}F^{\mu \nu}\right)^2$ and $\left(F_{\mu \nu} \tilde{F}^{\mu \nu}\right)^2$. But by dimensional analysis, they should appear at order $O(1/\beta^2)$ in a Taylor expansion, and as we already remarked, the three non-linear extensions exactly agree at this order, therefore we can take the same bound for the $\beta$ parameter in the three cases considered here with a good approximation. 

Therefore, considering $\sqrt{\beta} \ge 90 \, \text{GeV}$, we can obtain the estimated mass for the monopole configuration. Summarizing, considering these three different non-linear extensions we have:
\begin{subequations}
\begin{align}
	&E \approx %4,1 \, \text{TeV} + 51,87 \sqrt{\beta} \approx
	8.7 \,\, \text{TeV} \quad \left(\text{Logarithmic}\right),  \\
	&E \approx %4,1 \, \text{TeV} + 42,33 \sqrt{\beta} \approx
	7.9 \,\, \text{TeV} \quad \left(\text{Exponential}\right), \\
	&E \approx %4,1 \, \text{TeV} + 72,81 \sqrt{\beta} \approx
	11.6 \,\, \text{TeV} \quad \!\!\! \left(\text{Born-Infeld}\right). 
\end{align}
\end{subequations}
We remark here that our Logarithmic and Exponential non-linear extensions give a lower mass for the monopole solution than the one obtained with Born-Infeld, but unfortunately it is still above the threshold energy for pair production of this object at the present LHC. In the following, we will consider a simplified setup to discuss a lower bound for the monopole energy in each case of interest.

\section{Lower bounds for the monopole energy}

%To close the discussion, we will consider in the following a simplified setup to estimate a lower bound for the monopole energy. 
%First, we consider pure electromagnetism with the non-linear extensions directly in the $U(1)_{em}$. After, we consider the more realistic case of EW theory, but in the BPS limit, to estimate a more reliable lower bound for the monopole mass in our approach.

The energy functional for the EW monopole {\it ansatz} is
\begin{align}
E &=  \int_0^\infty dr \,  4 \pi r^2 \left[  \frac{\dot{\rho}^2}{2} + \frac{\rho^2 f^2}{4}\frac{1}{r^2} + \frac{\lambda}{8}(\rho^2 - \rho^2_0)^2 \right. + \nonumber \\  
&+ \left.    \frac{\dot{f}^2}{g^2} \frac{1}{r^2} + \frac{(f^2-1)^2}{2 g^2} \frac{1}{r^4}    +   \frac{1}{2 g'^2} \frac{1}{r^4}  \right]. 
\end{align}
Taking the so-called BPS limit \cite{Bogomolnyi, Prasad}, that is, taking the limit $\lambda \rightarrow 0$ but keeping the asymptotic condition $\rho \rightarrow \rho_0$, and also doing the improvement of the $U(1)_Y$ kinetic term for a non-linear version, we can rewrite the above expression as
\begin{align}\label{key}
	E &=  \int_0^\infty dr \,  4 \pi r^2 \left[ \left(\frac{\dot{\rho}}{\sqrt{2}} + \frac{(f^2-1)}{\sqrt{2} g r^2} \right)^2 + \left( \frac{\dot{f}}{g r } + \frac{f \rho}{ 2 r} \right)^2 \right. \nonumber \\
	& \left. - \frac{\dot{\rho} (f^2 - 1)}{g r^2} - \frac{\dot{f} f \rho}{g r^2} - \phi\left(\mathcal{F}\vert_{ansatz} \, , \, \mathcal{G}\vert_{ansatz} \right). \right]
\end{align}
The last term is the contribution of the non-linearly extended hypercharge kinetic term, it was already computed and is completely independent of $\rho $ and $f$, therefore we will omit it in our analysis. The terms in the first line are clearly non-negative, and therefore we can write a lower bound for the energy functional in this BPS limit,
\begin{align}\label{BPSorig}
	E \geq - \frac{4 \pi}{g}\int_0^\infty dr \,    \left[   \dot{\rho} (f^2 - 1) + \dot{f} f \rho %- \phi \left(\mathcal{F} , \mathcal{G} \right)\vert_{ansatz}
	\right].
\end{align}

To saturate the bound and obtain the configurations that minimize the energy in this setup, we need to consider configurations that solve the following equations:
\begin{align}\label{bogequations}
	&\dot{\rho}(r) + \frac{(f(r)^2 - 1)}{g r^2} = 0, \nonumber \\
	&\dot{f}(r) + \frac{g f(r) \rho(r)}{2} = 0.
\end{align} 
Interestingly, we would like to point out that if we didn't have a factor 2 in the denominator of the second equation, we would be able to find an analytical solution for these equations, as found by Bogomol'nyi \cite{Bogomolnyi}, Prasad and Sommerfield \cite{Prasad} for the 'tHooft-Polyakov monopole. Such analytical solution would be
\begin{align}\label{BPSanalitica}
	f(r) = \frac{g \rho_0 r}{\sinh(g \rho_0 r)}; \quad
	\rho(r) = \frac{\rho_0}{ \tanh(g \rho_0 r)} - \frac{1}{g r}.
\end{align}
Unfortunately, we were not able to find an analytic solution for our case, but even though, we can search for a numerical solution to these equations, only to be capable of estimating a lower bound to the monopole mass. We remark that once again, a factor 2 prevents us from obtaining a total derivative in the expression \ref{BPSorig}, resulting in an analytical and elegant result. 
%For this purpose, we can solve the equations \ref{bognumerica} numerically and use the profiles \ref{BPSanalitica} to fit the numerical solution. Doing this, we will be able to easily estimate a lower bound for the energy. We remark that we are not intending to obtain a precise value, but only a rough estimate for the minimum energy of such a configuration.
%The contribution that comes from the hypercharge kinetic term was already computed in the last Section and is completely independent of the functions $\rho $ and $f$, so we will not consider it here.  
Considering configurations that solve the above equations and therefore saturate the energy bound, we can rewrite the lower bound,
\begin{align}\label{lower bound cho}
	E \ge \int_0^\infty dr \, 4 \pi \left( \frac{\rho^2 f^2}{2} + \frac{(f^2 -1)^2}{g^2 r^2}     \right).
\end{align}

%Let us rewrite the equations \ref{bogequations} using adimensional variables, that is, consider, $r = x / m_W$, $ \rho = \hat{\rho} \rho_0$, where as we know, we can write $\rho_0 = \frac{2}{g} m_w $. Thus, we obtain,
%\begin{align}\label{bognumerica}
%	&\hat{\rho}'(x) + \frac{(f(x)^2 -1)}{2 x^2} = 0 \nonumber \\
%	&f'(x) + f(x) \hat{\rho}(x) = 0,
%\end{align}
%where the derivatives here are with respect to the adimensional variable $x = r \, m_W$. Therefore, we can rewrite the lower bound for the energy as,
%\begin{align}\label{key}
%		E \geq   \frac{8 \pi}{g^2} m_W \int_0^\infty dx \left[ \hat{\rho}(x) f(x)^2  + \frac{(f(x)^2 -1)^2}{2 x^2}   \right].
%\end{align}
%\begin{align}\label{key}
%	E \geq  - \frac{8 \pi}{g^2} m_W \int_0^\infty dx \left[ \hat{\rho}'(x) (f(x)^2 -1) + f'(x) f(x) \rho(x)   \right].
%\end{align}
%In fact, if we had an extra factor 2 in the second term in \ref{BPSorigi}, considering the asymptotic behavior of the fields given by \ref{asymptotic}, we would obtain as a bound for the energy the following expression,
%\begin{align}\label{key}
%	E \ge \frac{4 \pi }{g} \rho(\infty) = \frac{8 \pi }{g^2} m_W \approx 
%\end{align}
%Using the procedure described in the last paragraph, and using here $g = 0,65$, and $m_W = 80,37 $ GeV, we obtain:
%\begin{align}\label{key}
%	E \geq 2,01 \, \text{TeV}. 
%\end{align}
In the recent work \cite{Regcharge}, even though the authors used a different setup for the regularization of the monopole energy, when considering the BPS limit they obtained exactly the same expression for the equations that saturate the energy \ref{bogequations}, as well as the same integral for the energy lower bound \ref{lower bound cho} that comes as consequence, and the result obtained there for such integral is given by
\begin{align}\label{energyBPS}
	E \ge 2.98 \, \text{TeV}.
\end{align}
We remark for the sake of comparison that another BPS bound was already obtained in \cite{BPSCM}, giving a lower bound of $2.37 \, \text{TeV}$.
In our case, the result obtained above is a lower bound for the EW monopole energy ignoring not only the scalar potential contribution, but also the hypercharge kinetic term ones. Taking in consideration now the result obtained for the hypercharge sector in each of the non-linear extensions that we did before, we find an estimate for the more realistic setup of a EW monopole, 
\begin{subequations}
\begin{align}\label{key}
	&E \geq 7.6 \,\, \text{TeV} \quad \text{(Logarithmic)},  \\
	&E \geq 6.8 \,\, \text{TeV} \quad \text{(Exponential)},  \\
	&E \geq 10.5 \,\, \text{TeV} \quad \text{(Born-Infeld)}.
\end{align}   
\end{subequations}

Therefore, we conclude from our estimate that our non-linear extensions (i.e., Logarithmic and Exponential) give us a lower bound for the monopole mass that could be eventually found at the LHC, since the necessary energy to pair produce the monopole is nearby the present achievable energies. Therefore, even if our solutions have energy above the threshold for pair production at LHC, with these lower bounds we can have hope of some modification of our solution, that can give a monopole mass achievable at the present colliders.

\section{Concluding Remarks}

%In the following, we discuss the interesting case of a purely electromagnetic Dirac monopole.

To conclude our contribution, let us consider the simpler case of pure Electromagnetism and, following the same procedure as previously shown, let us find what the answer for a Dirac-like monopole would be. In fact, let us consider here the following Lagrangian:
\begin{align}\label{key}
	\mathcal{L}_{\text{EM}} = - \frac{1}{4} F_{\mu \nu} F^{\mu \nu} + f\left( \mathcal{F} , \mathcal{G} \right),
\end{align}
where now we will do the non-linear extension directly on the Electromagnetism, and we are defining here the invariants as $\mathcal{F} = \frac{1}{4} F_{\mu \nu} F^{\mu \nu} $ and $\mathcal{G} = \frac{1}{4} F_{\mu \nu} \tilde{F}^{\mu \nu} $.

Following the same steps as before, we will find
\begin{align}\label{key}
	\partial_\mu F^{\mu \nu} = \frac{F^{\mu \nu} \partial_\mu \partial_\mathcal{F} f + \tilde{F}^{\mu \nu} \partial_\mu \partial_\mathcal{G} f + \partial_\mu \tilde{F}^{\mu \nu} \partial_\mathcal{G} f }{1 - \partial_\mathcal{F} f}
\end{align}
Consider now the {\it ansatz} for a Dirac-like monopole,
\begin{align}\label{key}
	A_\mu = - \frac{1}{2 e} (1 - \cos\theta) \partial_\mu \varphi.
\end{align}
In the static regime we have $\partial_0 \equiv 0$, and therefore $A_0 = 0$, giving also $F_{0 i} = 0$, and $\tilde{F}_{\i j} = 0$ immediately.
Therefore, we will obtain the following energy functional:
\begin{align}\label{key}
	E = \int_0^\infty dr 4\pi r^2 \left[ \frac{1}{8 e^2 r^4} - f\left( \mathcal{F} = \frac{1}{8 e^2 r^4} , \mathcal{G} = 0 \right)   \right].
\end{align}

We already saw that the monopole {\it ansatz} gives a solution for the $U(1)$ equations of motion in the non-linear extensions that we considered here, and the same reasoning used before works for this case. Considering here the Logarithmic, Exponential and Born-Infeld Electrodynamics respectively, we obtain for the monopole energy,
\begin{align}\label{key}
	&E_{\text{Log}} = \frac{2^{1/4}\, \pi^2 }{2 e^{3/2}} \sqrt{\beta}, \nonumber \\
	&E_{\text{Exp}} = \frac{\pi\, \Gamma(1/4) }{3 \, 2^{1/4} e^{3/2}} \sqrt{\beta}, \nonumber \\
	&E_{\text{BI}} = \frac{3 \sqrt{\pi / 2} \, \Gamma(-3/4)^2}{16 e^{3/2}} \sqrt{\beta}.
\end{align}
Taking $e = 0.303$ and considering the bound obtained in \cite{Ellis2} that gives $\sqrt{\beta} \ge \, 100 \text{GeV}$ for the nonlinear extension directly in the Electromagnetism, we obtain
\begin{subequations}
\begin{align}\label{key}
E_{\text{Log}} &\approx 2.3 \, \text{TeV},  \\
E_{\text{Exp}} &\approx 1.9 \, \text{TeV},  \\
E_{\text{BI}}  &\approx 3.3 \, \text{TeV}.
\end{align}
\end{subequations}
Therefore, we can see what is the mass of a Dirac monopole if we consider only the Electromagnetism with a non-linear extension. This is a simplified scenario, but even though, it can give us a lower bound for the monopole mass, in a scale achievable at LHC. 
%In the next Section, we will do few considerations of what would be a more reliable lower bound for the EW monopole energy.

The electroweak theory is extremely successful, but there still remains an important unanswered question of topological nature. In fact, even if it has never been observed, it can be shown that it admits EW monopole solutions, with classical infinite energy, rendering, therefore, impossible to predict its mass. We are presenting here a regularization for the EW monopole energy obtained by extending the hypercharge sector to a non-linear version based on Logarithmic and Exponential versions of Electrodynamics. Furthermore, we identified the constraints that a more general non-linear extension should obey to yield finite energy solutions. We have also worked out an estimate of the monopole mass in each non-linear scenario here contemplated; the results are compared with the result already known for the BI extension. We conclude that, in the cases we investigate, our monopole solutions are lighter than the known BI solutions, but, unfortunately, our masses remain still out of reach for the current colliders. We estimate the lower bound for the monopole energy in our approach and conclude that it is possible to suitably modify our solution to have an energy accessible at LHC.

%%%%%%%%%%%%%%%%%%%%%%%%%%%%%%%%%%%%%%%%%%%%%%%%%%%%%%%%%%%%%

\begin{acknowledgments}
This work was partially funded by Brazilian National Council for Scientific and Technological Development (CNPq). PDF would like to thank W. B. de Lima, G. P. de Brito and P. C. Malta for useful discussions.
	
\end{acknowledgments}

%%%%%%%%%%%%%%%%%%%%%%%%%%%%%%%%%%%%%%%%%%%%%%%%%%%%%%%%%%%%%%

\end{document}